\documentclass[twocolumn,prb,showpacs]{revtex4}

\usepackage{amsmath}
\usepackage{amssymb}
\usepackage{epsfig}

\newcommand{\tr}{\text{Tr}}

\begin{document}
\title{Kekule-distortion-induced Exciton instability in graphene}

\author{Raoul Dillenschneider}
\email[E-mail address : ]{raoul.dillenschneider@physik.uni-augsburg.de}
\affiliation{Department of Physics, University of Augsburg, D-86135 Augsburg, 
Germany}

\date{\today}

\begin{abstract}
Effects of a Kekule distortion on exciton instability in single-layer 
graphene are discussed.
In the framework of quantum electrodynamics the mass of the electron 
generated dynamically is worked out using a Schwinger-Dyson equation. 
For homogeneous lattice distortion it is shown that the generated mass
is independent of the amplitude of the lattice distortion at the one-loop 
approximation. Formation of excitons induced by the homogeneous Kekule
distortion could appear only through direct dependence of the lattice 
distortion.
\end{abstract}

\pacs{71.30.+h,71.35.-y,11.10.Kk,81.05.Uw}

\maketitle

\section{Introduction}

Recent experimental accessibility of graphene
\cite{graphene-QH_P1,graphene-QH_P2,graphene-QH_P3}
has drawn much interest on this material which present a wide variety 
of interesting properties \cite{Geim_Novoselov_2007,McCannFalko,AbergelFalko,Baskaran,Peres}.
Most of the properties of the graphene arises from the peculiar energy 
spectrum near the so-called Dirac nodal points 
\cite{Semenoff,Novoselov_Nature438}.

Recent work on exciton instability in graphene monolayers
is based on the Dirac Hamiltonian description
\cite{Khveshchenko,LealKhveshchenko,KhveshchenkoShively}. The exciton gap is
derived and solved through a self-consistent equation similar to the
one appearing in the chiral symmetry breaking phenomenon
\cite{Appelquist1}. It was shown that an exciton can be formed under
a strong long-ranged particle-hole
interaction\cite{LealKhveshchenko,Gusynin}. Exciton can also be formed in a
single-layer graphene through the mechanism of magnetic catalysis of
dynamical mass generation as pointed out in
\cite{GusyninPRL}. This work showed that the magnetic
catalysis can induce exciton condensation even for weak
particle-hole coupling \cite{Gusynin}. These results are obtained
in the framework of quantum electrodynamics $QED$ deduced from the
linear energy spectrum of the graphene monolayer.

Exciton instability in graphene bilayer systems have been studied in the
case of a short-ranged Coulomb interaction and a finite voltage
difference between the layers \cite{JHHanDR}. Self-consistent exciton gap
equations are derived in the framework of Hartree-Fock approximation
and it is shown that a critical strength of the Coulomb interaction exists 
for the formation of excitons. The critical strength depends on the amount 
of voltage difference between the layers and on the inter-layer hopping 
parameter. The voltage difference drives a gap in the energy spectrum
of the graphene bilayer \cite{Castro,OOstinga} and combined to a
strong Coulomb interaction leads to an exciton instability.

Similarly a gap can open in the energy spectrum of graphene monolayers
by means of a lattice distortion such as the so called Kekule distortion 
\cite{Chamon,HouChamon,JackiwPi}. A weakly screened Coulomb interaction
and a gap in the linear energy spectrum of the graphene monolayer are
favorable elements for the formation of excitons.

We focus our attention here on the consequences of a Kekule distortion
on dynamical mass generation for electrons in the graphene monolayer.
Indeed quasiparticle in 2D systems can acquire mass through dynamical
interaction between electrons and holes leading to an increase of 
the energy gap \cite{DoreyMavromatos,Appelquist1,Appelquist2}.
We address the question : can a Kekule distortion affect the dynamical
mass generation mechanism in such a way that the resulting energy gap 
favors the formation of excitons ? We show that in the specific case
of an homogeneous Kekule distortion and at the one-loop approximation 
the dynamical mass generation is unaffected by the lattice distortion. 
We show that the gap in the energy spectrum is a sum of two independent 
contributions : one induced by the homogeneous Kekule distortion 
and one induced by dynamical mass generation. 
Consequently exciton instability can only be formed 
through direct dependence of the amplitude of the lattice distortion
and the Kekule-independent dynamical mass of the electrons.
There is no amplification effects of the homogeneous Kekule distortion
by the mechanism of dynamical mass generation.

The outline of the paper is the following. In section \ref{Section1}
we recall the construction of the quantum electrodynamic action
of the graphene monolayer in presence of a Kekule distortion.
In section \ref{Section2} we present the calculation of the dynamical mass 
term using the Schwinger-Dyson equation of the electron. 
Section \ref{Section3} summarizes and discusses the present work.

\section{$QED_3$ action of the graphene monolayer with a Kekule distortion
\label{Section1}}

\begin{figure}[h]
\center
\epsfig{file=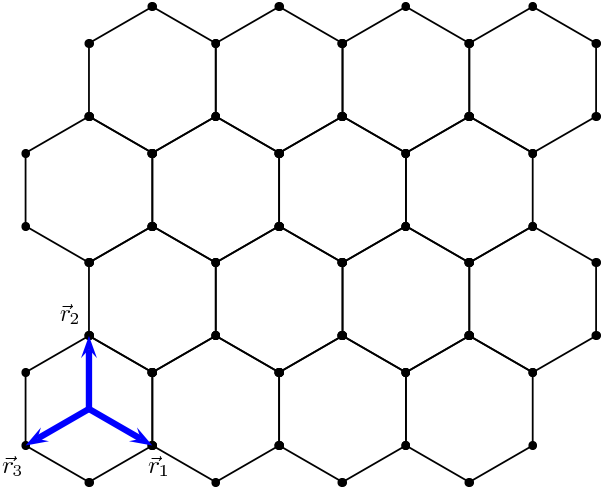,width=8cm}
\caption{Representation of the graphene monolayer with the nearest-neighbour 
vectors $r_\alpha$ connecting the two sublattices $\Lambda_A$ and $\Lambda_B$.}
\label{Fig1}
\end{figure}

A graphene monolayer is a honeycomb array of atoms of carbon 
as depicted in Fig. \ref{Fig1}. In a monolayer the electrons 
can hop between nearest-neighbour carbon atoms through $\pi$-orbitals 
with energy $t$.
The unit cell of the graphene monolayer is composed of two type of 
carbon atoms and we denote them by $A$ and $B$. The whole lattice is
devided in to two sublattices, the $\Lambda_A$-type and the $\Lambda_B$-type.
The Hamiltonian describing the graphene monolayer reads

\begin{eqnarray}
H_0 &=&
-t \sum_{r \in \Lambda_{A}} \sum_{\alpha = 1}^3 a^\dagger_r b_{r+r_\alpha} 
+ h.c.,
\label{Eq1}
\end{eqnarray}

\noindent
where $a$ and $b$ stand for the $\pi$-electron creation and anihilation 
fermionic operator on the atoms of type $A$ and $B$ respectively. 
The vector $r_\alpha$ connects the two sublattices 
$\Lambda_A$ and $\Lambda_B$ and is given by $r_\alpha = a e_\alpha$ with 
$a$ the lattice parameter and $e_1 = (1,0)$, $e_2 = (-1/2,\sqrt{3}/2) $ 
and $e_3 = (-1/2,-\sqrt{3}/2)$ are unit vectors. 
In the following the lattice parameter $a$ 
will be set equal to one and plays the role of the unit of length.

The diagonalisation of the Hamiltonian \eqref{Eq1} leads to the kinetic energy 
$\varepsilon_k = t|\sum_{\alpha=1}^3  e^{i k.e_\alpha}|$, where
$e_\alpha$'s are the nearest-neighbour vectors of the graphene
monolayer. The kinetic energy vanishes at the two independent nodal points 
$K_{+(-)}$ which are chosen as $K_{+} = \left(0,\frac{4\pi}{3\sqrt{3}}\right)$
and $K_{-} = -K_{+} $ in the Brillouin zone.
At low energy the bare Hamiltonian $H_0$ of the $\pi$-electron can be 
rewritten by considering the energetic contributions arround the nodal 
points and reads \cite{Semenoff}

\begin{eqnarray*}
H_0 &=& 
-\Bigg\{
\sum_r v_F u_a^\dagger(r) (2 \partial_{\bar{z}}) u_b(r)
\notag \\
&&
+\sum_r v_F v_a^\dagger(r) (-2 \partial_{z}) v_b(r)
+h.c.
\Bigg\},
\end{eqnarray*}

\noindent
where we made use of the notations in the complex plan $z = x + iy$ and 
$\partial_z = \frac{1}{2}\left(\partial_x - i \partial_y \right)$.
The velocity $v_F = \frac{3}{2}t a/\hbar$ where we set $\hbar = a =1$ 
(the kinetic energy becomes $\varepsilon_p = \hbar v_F |p|$).
The amplitude $u_a(b)$ and $v_a(b)$ are smooth functional operators and 
are connected to the creation and annihilation operator $a$ and $b$ through 
the Fourier components $u_a(p) = a_{p+K_{+}},u_b(p) = b_{p+K_{+}}$ 
and $v_a(p) = a_{p+K_{-}},v_b(p) = b_{p+K_{-}}$ where $p$ is the wave vector.
The Hamiltonian $H_0$ describes the hopping of the $\pi$-electrons in an 
undistorted lattice.

A lattice distortion can be modeled by a variation of the hopping parameter 
$\delta t_{r,\alpha}$ which depends on the position in the lattice $\vec{r}$ 
and on the direction $e_\alpha$. The corresponding Hamiltonian for the 
graphene monolayer reads

\begin{eqnarray}
H_K = 
-\sum_{r \in \Lambda_{A}} \sum_{\alpha = 1}^3 \delta t_{r,\alpha} 
a^\dagger_r b_{r+b_\alpha} + h.c.
\end{eqnarray}

\noindent
In the following we consider the Kekule distortion for which the 
bonds between the carbon atoms are arranged similar
to the benzene molecule \cite{Chamon}.
The lattice is then pictured by an alternation of long and short bonds between 
the carbon atoms of types $A$ and $B$ \cite{Viet}. 
The Kekule distortion is modeled by the hopping parameter
$\delta t_{r,\alpha}$ and reads
$
\delta t_{r,\alpha} = \frac{1}{3} 
\left( 
\Delta(r) e^{i K_{+} r_\alpha} e^{i G.r}
+ 
\bar{\Delta}(r) e^{i K_{-} r_\alpha} e^{-i G.r}
\right)
$,
where $\Delta(r)$ stands for the amplitude of the Kekule distortion and
$K_{+(-)}$ are the nodal points in the Brillouin zone. The vector 
$G = K_{+} - K_{-}$ connects the two independent Dirac cones located at the
nodal points $K_{+(-)}$.
Focusing on the low-energy contributions of the Kekule distortion around the
nodal points the Hamiltonian $H_K$ is given by

\begin{eqnarray*}
H_K = 
-\sum_{r \in \Lambda_A} 
\left[
\Delta(r) u_a^\dagger (r) v_b(r) + \bar{\Delta}(r) v_a^\dagger (r) u_b(r)
\right]
+ h.c.
\end{eqnarray*}

\noindent
The Hamiltonian describing the $\pi$-electrons in the graphene single-layer
with a Kekule distortion is then the sum of the bare and Kekule 
Hamiltonians $H = H_0 + H_K$. Using
the fermionic spinor $\psi^\dagger  = \left[u_b(r) u_a(r) v_a(r) v_b(r) 
\right]^\dagger$ the Hamiltonian $H$ can be rewritten in the 
quantum electrodynamic framework and reads

\begin{eqnarray*}
H = \int d^2 \vec{r} \bar{\psi}(r) \left[v_F \gamma_k \partial_k 
+ \widetilde{\widetilde{\Delta}}(r) \right] \psi(r),
\end{eqnarray*}

\noindent
where the gamma matrices are defined by $\gamma_0 = \tau_3 \otimes \tau_3$,
$\gamma_1 = v_F \tau_1 \otimes \tau_3$ and 
$\gamma_2 = v_F \tau_2 \otimes \tau_3$, $\tau_{\{0,1,2\}}$ are the
Pauli matrices. The element $\widetilde{\widetilde{\Delta}}(r)$
is a $4 \times 4$ matrices related to the Kekule distortion amplitude
$\Delta(r)$ by the relation 
$
\widetilde{\widetilde{\Delta}}(r) = 
\left( 
\begin{array}{cc}
0 & \Delta(r) \tau_3 \\
- \bar{\Delta}(r) \tau_3 & 0
\end{array}
\right)
$.
The aim of the present work being to characterize the behaviour of the 
dynamical mass generation for graphene monolayer in presence of an 
homogeneous Kekule distortion we reduce our study to the
case $\Delta(r)=\Delta_0$.
Finally for a system at temperature $T = 1/\beta$ 
the action of the graphene monolayer is given by

\begin{eqnarray}
S_{el} =
\int_0^\beta d\tau \int d^2 \vec{r} \bar{\psi}(r,\tau)
\left[
\gamma^\mu \partial_\mu + \widetilde{\widetilde{\Delta}}_0
\right]
\psi(r,\tau).
\label{Eq3}
\end{eqnarray}

The presence of the electromagnetic field surrounding the graphene monolayer
is also to be considered. Indeed the graphene monolayer is embedded in an 
electromagnetic field which spans the whole 3D space. However the electrons 
are confined in the 2D space delimited by the carbon atoms. 
Hence the density of charge and current verify 
$\rho(x,y,z) = \delta(z) \rho_{2D}(x,y)$ and 
$\vec{J}(x,y,z) = \delta(z)\vec{J}_{2D}(x,y)$. The electromagnetic field 
vector $\vec{a}$ and scalar $\phi$ potentials are related to the
Green functions of the laplacian $2\sqrt{-\vec{\nabla}^2}$ rather 
than $\vec{\nabla}^2$ for the three-dimensionnal electromagnetic field 
\cite{DoreyMavromatos,GusyninPRD}. 
To describe the electromagnetic field embedding the graphene monolayer 
it is convenient to use the following Euclidean $QED_3$ action

\begin{eqnarray}
S_{e.m.} 
= -\int d^3 x \frac{1}{2 \sqrt{-\partial^2}} f^{\mu \nu} f_{\mu \nu},
\label{Eq4}
\end{eqnarray}

\noindent
where $f_{\mu \nu} = \partial_\mu a_\nu - \partial_\nu a_\mu$ is the 
electromagnetic field tensor derived from the three-dimensionnal
vector potential $a_\mu$. The Fourier transform of action \eqref{Eq4}
leads to the bare photon propagator $\Delta_{\mu \nu}^{(0)} = \left(
\delta_{\mu \nu} - q_\mu q_\nu \right)/(2 |q|)$ in Euclidean space.
For finite temperature the imaginary-time component $q_0$ of the 
(2+1)-dimensionnal wave-vector $q=(q_0,q_1,q_2)$ is given by the
bosonic Matsubara frequency $q_0 = 2 \pi n /\beta$ ($n$ is an integer and
takes its values in the range $]- \infty ,\infty [$).

Finally the full quantum electrodynamic action describing the graphene
monolayer with an homogeneous Kekule distortion and embedded in a 3D
electromagnetic field reads $S = S_{e.m.} + S_{el}$

\begin{eqnarray}
S
&=& 
\int d^3 x \frac{-1}{2 \sqrt{-\partial^2}} f_{\mu \nu} f^{\mu \nu}
\notag \\
&&
+ \int d^3 x \bar{\psi} \left[
\gamma^\mu \left( 
\partial_\mu - i g a_\mu
\right) + \widetilde{\widetilde{\Delta}}_0 \right]\psi .
\label{Eq5}
\end{eqnarray}

It is known that 2D systems described by a quantum electrodynamic
action like \eqref{Eq5} experience a dynamical mass generation for the
fermionic field $\psi$ in interaction with an $U(1)$ gauge field. 
Appelquist et al. \cite{Appelquist1,Appelquist2} showed that at zero 
temperature the originally massless fermions can acquire a dynamically
generated mass when the number $N$ of fermion flavors is lower than the 
critical value $N_c = 32/\pi^2$. Later Maris \cite{Maris} confirmed the 
existence of a critical value $N_c \simeq 3.3$ below which the dynamical 
mass can be generated. Since we consider only spin-$1/2$ systems, 
$N=2$ and hence $N<N_c$.
At finite temperature Dorey and Mavromatos \cite{DoreyMavromatos} and Lee
\cite{Lee-98} showed that the dynamically generated mass vanishes at a 
temperature $T$ larger than the critical one $T_c$.
More recents works have been performed on dynamical mass generation in 
graphene monolayer \cite{Khveshchenko,LealKhveshchenko}.
However the question now arise about the effects of a lattice distortion 
like the Kekule distortion. How does the mass generated dynamically
in presence of a Kekule distortion behave?
We concentrate on the effects of an homogeneous Kekule distortion 
and for small temperature $T \rightarrow 0$.

\section{Dynamical mass generation \label{Section2}}

\subsection{The photon propagator at finite temperature}

Integrating over the fermion fields $\psi$ leads to a pure gauge Lagrangian
$\mathcal{L}_a = \frac{1}{2} a_\mu \Delta_{\mu \nu}^{-1} a_\nu$ where
$\Delta_{\mu \nu}$ is the dressed photon propagator from which we shall  
extract an effective interaction potential between two fermion
and derive the dynamical mass of the fermions.

The finite temperature photon propagator in Euclidean space 
verifies the Dyson equation \cite{Das,JR3}

\begin{eqnarray*}
\Delta_{\mu \nu}^{-1} &=& {\Delta_{\mu \nu}^{(0)}}^{-1} + \Pi_{\mu \nu},
\label{Eq6}
\end{eqnarray*}

\noindent
where the bare photon propagator $\Delta_{\mu \nu}^{(0)}$ is derived from
the action of the bare electromagnetic field \eqref{Eq4} and
the polarisation function is obtained from the integration over the fermionic
field $\psi$ in the action \eqref{Eq5}.

For the computation of the dynamically generated mass it is enough
to consider the static temporal dressed photon propagator component
$\Delta_{00}(q^0=0,\vec{q})$ for which the bare photon
propagator component reads ${\Delta_{00}^{(0)}(q^0=0,\vec{q})} = 1/2|q|$.
The detailed calculation of the static temporal component of the 
polarisation function $\Pi_{00}(q^0=0,\vec{q})$ is given in appendix 
\ref{AppendixA} and reads

\begin{eqnarray}
\Pi^{00}(q^0=0,\vec{q}) &=& 
\int_0^1 dx
\left(
\frac{\alpha}{2\pi v_F^2 \beta}
\right)
\Bigg\{
\ln \left(2 \cosh{(\pi \Theta_q)} \right)
\notag \\
&&
- \left( \frac{\beta |\Delta_0|}{2} \right)^2
\frac{\tanh{(\pi \Theta_q)}}{(\pi \Theta_q)}
\Bigg\},
\label{Eq7}
\end{eqnarray}

\noindent
where $\beta$ is the inverse temperature and 
$\Theta_q = \left(\frac{\beta}{2\pi} \right)
\sqrt{x(1-x)v_F^2 \vec{q}^2 + |\Delta_0|^2}$. The coupling parameter
$\alpha$ is related to the number of fermion flavor $N=2$ and to the
electron charge $g$, $\alpha = 4 g^2 N$.

For very small temperature $T \rightarrow 0$ the polarization function
\eqref{Eq7} becomes 

\begin{eqnarray*}
\Pi^{00}(q^0=0,\vec{q}) &=& 
\left( \frac{\alpha}{8\pi v_F^2} \right)
\Bigg[
|\Delta_0| 
\notag \\
&&
+ \frac{\left( v_F q \right)^2 - 4 |\Delta_0|^2}{2 v_F |q|}
\arctan \left( \frac{v_F q}{2 |\Delta_0|} \right)
\Bigg].
\end{eqnarray*}

\noindent
The process of dynamical mass generation is dominated
by mechanisms at large wave vectors $\vec{q}$. Indeed for 
$q \gg 2 |\Delta_0|/v_F$ the polarisation function is asymptoticaly equal to 
$\frac{\alpha}{16 \pi v_F} |q|$ which confirms the fact that the dynamics
between electrons and holes are dominated by large wave vectors \cite{Gusynin}.

\subsection{The electron self-energy}

We now derives the electron self-energy which is also the dynamical mass. 
The Schwinger-Dyson equation for the electron propagator at finite 
temperature reads

\begin{eqnarray}
G^{-1}(k) &=& {G^{(0)}}^{-1}(k) 
\notag \\
&-& \frac{g}{\beta} \underset{\widetilde{\omega}_{F,n}}
{\sum} \int \frac{d^2 \vec{P}}{(2 \pi)^2} \gamma_\mu G(p) \Delta_{\mu \nu}
(k-p) \Gamma_\nu,
\notag \\
\label{SchwingerDyson}
\end{eqnarray}

\noindent
where $p=(p_0=\widetilde{\omega}_{F,n},\vec{P})$,
$G$ is the dressed electron propagator, $\Gamma_\nu$ the electron-photon
vertex which will be approximated here by its bare value $g \gamma_\nu$ and
$\Delta_{\mu \nu}$ is the dressed photon propagator.
The second term in \eqref{SchwingerDyson} is the fermion self-energy $\Sigma$,
($G^{-1} = {G^{(0)}}^{-1} - \Sigma$).
Performing the trace over the $\gamma$ matrices and working out
the sum over the fermionic Matsubara frequencies $\widetilde{\omega}_{F,n}$ 
in equation \eqref{SchwingerDyson} leads to a self-consistent equation for 
the self-energy of the form

\begin{eqnarray}
\Sigma(\vec{k}) = g^2 \int \frac{d^2\vec{p}}{(2\pi)^2}
\Delta_{00}(0,\vec{k}-\vec{p}) \frac{\Sigma(\vec{p})}{2 \varepsilon_{\vec{p}}}
\tanh{\left(\frac{\beta \varepsilon_{\vec{p}}}{2} \right)},
\label{Eq8}
\end{eqnarray}

\noindent 
where $\varepsilon_{\vec{p}} = \sqrt{\varepsilon^{(0) 2}_{\vec{p}} 
+ \Sigma(p)^2  }$ and $\varepsilon^{(0) 2}_{\vec{p}} 
= v_F^2 \vec{p}^2 + |\Delta_0|^2$ are respectively the energy spectrum
of the graphene monolayer with and without correction of the one-loop
approximation near the nodal points $K_{+(-)}$. 
In the limit $q \gg 2 |\Delta_0|/v_F$
the dressed photon propagator is given by 
$\Delta_{00}(0,\vec{q})=\left[(2+\alpha/(16 \pi v_F))|q|\right]^{-1}$.
The angular integration in equation \eqref{Eq8} is achieved using the
approximation $f(|\vec{p}-\vec{k}|) = \theta(p-k) f(|p|)
+ \theta(k-p) f(|k|)$ where $f$ is a function which depends on the
absolute value of its arguments \cite{Gusynin,KhveshchenkoShively}.
Hence the self-consistent equation \eqref{Eq8} reduces to

\begin{eqnarray}
\Sigma(k) &=& C \Bigg[
\int_0^k dp \frac{p \Sigma(p)}{k \sqrt{p^2 + {(\widetilde{\Delta}(p)/v_F)}^2} }
\notag \\
&&
+
\int_k^\Lambda dp \frac{\Sigma(p)}{\sqrt{p^2 
+ {(\widetilde{\Delta}(p)/v_F)}^2}}
\Bigg],
\label{Eq9}
\end{eqnarray}

\noindent
where we introduced the constant $C= 4g^2 /\left(32 \pi v_F + \alpha \right)$, 
the function $\widetilde{\Delta}(p)=\sqrt{|\Delta_0|^2 + \Sigma(p)^2}$ and
the ultraviolet cutoff $\Lambda$. The electron self-energy is
assumed to be a small correction to the electron propagator.
It follows that for $k \gg |\Delta_0|/v_F$ the denominator of the
second term in equation \eqref{Eq9} is approximately equal to $p$.
In the first term of equation \eqref{Eq9} the function $\widetilde{\Delta}(p)$
plays the role of a cutoff for $p \rightarrow 0$ and
equation \eqref{Eq9} can be rewritten
$
\Sigma(k) =
C 
\left[
\int_{\widetilde{\Delta}(p=0)/v_F}^k dp \frac{\Sigma(p)}{k}
+
\int_k^\Lambda dp \frac{\Sigma(p)}{p}
\right]
$.
A derivation of this equation up to the second order with
respect to the wave vector $k$ of the electron self-energy
leads to the differential equation

\begin{eqnarray}
k^2 \Sigma^{''}(k) + 2 k \Sigma^{'}(k) + C \Sigma(k) =0
\label{Eq10},
\end{eqnarray}

\noindent
for which the infrared and ultraviolet boundary conditions are given
respectively by

\begin{eqnarray}
k^2 \Sigma^{'}(k) {\Big\vert}_{k=\widetilde{\Delta}/v_F} &=& 0, \text{ and}
\label{Eq11}
 \\
\left(k \Sigma^{'}(k) + \Sigma(k) \right){\Big\vert}_{k=\Lambda} &=& 0.
\label{Eq12}
\end{eqnarray}

The differential equation \eqref{Eq10} admits a solution which verifies
also the boundary conditions \eqref{Eq11} and \eqref{Eq12} and reads
\cite{Appelquist2,Gusynin,KhveshchenkoShively}

\begin{eqnarray}
\Sigma(k) &=&
\frac{\widetilde{\Delta}^{3/2}}{\sin(\delta) \sqrt{v_F k}}
\sin \left(
\frac{\tan(\delta)}{2}
\ln \left( \frac{v_F k}{\widetilde{\Delta}}\right) + \delta
\right),
\notag \\
\label{Eq13}
\end{eqnarray}

\noindent
where the phase $\delta$ is equal to $\arctan \sqrt{4C-1}$.
In order to verify the boundary condition \eqref{Eq12} the amplitude
$\widetilde{\Delta}(p=0)$ has to verify the following relation
$
\widetilde{\Delta} = (v_F \Lambda) \exp{\left[- 4 \delta /\sqrt{4C-1} \right]}
$.

The solution of the electron self-energy has been derived for
wave vectors larger than the cutoff $|\Delta_0|/v_F$ and shows
to be independent of the amplitude $\Delta(r)=\Delta_0$ 
of the homogeneous Kekule distortion.
The independence of $\Sigma$ on the amplitude $\Delta_0$ implies that
the dynamical mass generation mechanism cannot be controlled by an
homogeneous Kekule distortion.

It was shown elsewhere in the context of graphene bilayer
that a gap opening in the energy spectrum can lead to an exciton
instability under the specific amplitude of a short-ranged Coulomb interaction
\cite{JHHanDR}. Here a similar excitonic instability could appear since
the graphene monolayer spectrum sees a gap open directly induced
by the Kekule distortion. 
Moreover the gap is enlarged by formation of a mass which is dynamically 
generated. 
An exciton instability could take place 
\cite{KhveshchenkoShively}. We address the question whether  
a Kekule distortion can amplify the mechanism of dynamical mass
generation and consequently affect the exciton instability.
The solution \eqref{Eq13} for the electron self-energy 
$\Sigma$ shows that the dynamically generated mass of the electron 
is independent of the homogeneous Kekule distortion. 
However the mass of the electron given by $m(k) = \sqrt{|\Delta_0|^2 +
\Sigma(k)^2} = \langle \bar{\Psi}(k) \Psi(k) \rangle$ shows that the
exciton instability resulting from the electron-hole interaction
is directly dependent on the homogeneous Kekule distortion but not through 
the mechanism of dynamical mass generation.

\section{Conclusions \label{Section3}}

The electron self-energy for graphene monolayer with an homogeneous Kekule 
distortion has been derived in the framework of quantum electrodynamics. 
It has been shown that the dynamical mass generation of the electrons
resulting from the electron-hole interaction in the graphene
monolayer is independent of the amplitude of the homogeneous Kekule distortion
for a one-loop approximation.
Such an independence of the dynamically generated mass provides an insight 
on the effects implied by lattice distortion on the energy spectrum 
of graphene monolayer.

The gap in the energy spectrum of the graphene monolayer is related
to the mass of the electron which is equal to $m(k) = \sqrt{|\Delta_0|^2 +
\Sigma(k)^2} = \langle \bar{\Psi}(k) \Psi(k) \rangle$. The Kekule
distortion control directly the gap by means of the amplitude $|\Delta_0|$ 
without any amplification through the dynamical mass $\Sigma$
(due to the independence mentioned previously).
Consequently exciton instability can be formed by direct relation to the 
homogeneous Kekule distortion and by the Kekule-independent mechanism
of dynamical mass generation.

So far we tried to shed light on the effects induced by lattice
distortions such as an homogeneous Kekule distortion $\Delta(r)=\Delta_0$.
The present study does not admit any conclusion on the independence of 
the dynamical mass $m(k)$ in presence of an inhomogeneous Kekule 
distortion $\Delta(r) \neq const$. It is expected that inhomogeneous
lattice distortion affects significantly the energy spectrum of
graphene monolayer. As a consequence exciton instability could possibly be
amplified through lattice distortion via non-trivial dynamical mass generation.

\acknowledgments{The author is grateful to Jean Richert for having 
attracted his attention to articles about Kekule distortion 
and for enlightening comments on the present work. 
The author would like to thank Professor Jung Hoon Han for 
enriching discussions.}

\appendix
\section{Derivation of the polarisation function \label{AppendixA}}

The polarisation function $\Pi_{\mu \nu}$ is obtained by integrating
the $QED_3$ action \eqref{Eq5} over the fermionic field $\psi$ and reads

\begin{eqnarray*}
\Pi^{\mu \nu}(q) = \frac{g^2}{\beta} \sum_{\sigma = \pm}
\sum_{\omega_f} \int \frac{d^2 \vec{k}}{(2\pi)^2}
\tr \left[
G_0(k)
.
\gamma^\mu
.
G_0(k+q)
.
\gamma^\nu
\right]
\end{eqnarray*}

\noindent
where $G_0(k) = \frac{\gamma^\rho k_\rho + i\widetilde{\widetilde{\Delta}}_0}
{k^2 + |\Delta_0|^2 }$ is the electron Green function and the
trace operator $\tr$ runs over the space of the gamma matrices.
Focusing only on the static temporal component of the polarisation 
$\Pi_{00}(q^0=0,\vec{q})$ and using the following relations
$
\tr \left[ 
i \widetilde{\widetilde{\Delta}}_0 \gamma^0 
i \widetilde{\widetilde{\Delta}}_0 \gamma^0
\right]
= -4 |\Delta_0|^2
$
and
$
\tr \left[
\gamma^\rho k_\rho .\gamma^0 \gamma^\eta (k_\eta + q_\eta) \gamma^0
\right]
=
4 g^{(M)}_{\rho \eta} k_\rho (k_\eta + q_\eta)
$
\noindent
where we defined the metric tensor 
$g^{(M)}_{\rho \eta} = diag(1,-v_F^2,-v_F^2)$, the polarisation function
can be reduced to

\begin{eqnarray}
&&\Pi^{00}(q) = \frac{g^2}{\beta} \sum_{\sigma = \pm}
\sum_{\omega_f} \int \frac{d^2 \vec{k}}{(2\pi)^2}
\Bigg(
\frac{1}{k^2 + |\Delta_0|^2}
\notag \\
&&
 \times
\frac{1}{(k+q)^2 + |\Delta_0|^2}
\Bigg)
\times
\Bigg\{
4 g^{(M)}_{\rho \eta} k_\rho (k_\eta + q_\eta)
-4 |\Delta_0|^2
\Bigg\}
\notag \\
\label{EqA2}
\end{eqnarray}

The computation of the polarisation function can be simplified using the
Feynmann identity $\frac{1}{ab} = \int_0^1 dx \frac{1}{\left(a x 
+ (1-x)b \right)^2}$,
 applying the change of variable $k \rightarrow k^{'} - x q$ 
(in other words $\omega_f^{'} = \frac{2\pi}{\beta} \left(n+1/2 \right)$ 
and $\vec{k}^{'} = \vec{k} + x \vec{q}$) and performing the sum
over the fermionic Matsubara frequencies $\omega^{'}_f$ one gets

\begin{eqnarray*}
&&
\Pi^{00}(q^0=0,\vec{q}) =
\notag \\
&&
\frac{\alpha}{\beta}
\int \frac{d^2 \vec{k}^{'}}{(2\pi)^2}
\int_0^1 dx
\Bigg\{
S_1 - 2\left(v_F^2 \vec{k}^{'2} + |\Delta_0|^2\right) S_2
\Bigg\}
\end{eqnarray*}

\noindent
We define by $S_1$ and $S_2$ the sums over the Matsubara frequencies 
given by the following relations \cite{Gradshteyn}

\begin{eqnarray*}
S_1 &\equiv& \sum_{n=-\infty}^{\infty} 
\frac{1}
{\left[{\omega^{'}}^2 + v_F^2 \vec{k}^{'2} + x(1-x)v_F^2 \vec{q}^2 
+ |\Delta_0|^2 \right]}
 \notag \\
&=&
\frac{\beta^2}{4\pi Y} \tanh (\pi Y)
 \\
S_2 &\equiv& \sum_{n=-\infty}^{\infty} \frac{1}
{\left[{\omega^{'}}^2 + v_F^2 \vec{k}^{'2} 
+ x(1-x)v_F^2 \vec{q}^2 + |\Delta_0|^2 \right]}
\notag \\
&=& -\frac{\beta^2}{8 \pi^2} \frac{1}{Y} \frac{\partial S_1}{\partial Y}
\end{eqnarray*}

\noindent
where $\omega^{'} = \frac{2\pi}{\beta}(n+1/2)$. The integration over the wave
vector $k^{'}$ can be performed through the change of variable
$Y = \frac{\beta}{2\pi} \sqrt{v_F^2 \vec{k}^{'2} 
+ x(1-x)v_F^2 \vec{q}^2 + |\Delta_0|^2}$ and finally the polarisation function
reads

{\small
\begin{eqnarray*}
\Pi^{00}(q^0=0,\vec{q}) &=&
\frac{\alpha}{\beta}
\int_0^1 dx
\lim_{\Lambda \rightarrow \infty} \int_{\Theta_q}^\Lambda
\frac{2\pi}{(v_F \beta)^2}
Y dY
\left[
S_1 + Y \frac{\partial S_1}{\partial Y}
\right]
\notag \\
&&
-
\frac{\alpha}{\beta}
\int_0^1 dx
\lim_{\Lambda \rightarrow \infty} \int_{\Theta_q}^\Lambda
\frac{dY}{2\pi}
x(1-x)\vec{q}^2 \frac{\partial S_1}{\partial Y}
  \notag \\
&=&
\int_0^1 dx
\left(
\frac{\alpha}{2\pi v_F^2 \beta}
\right)
\Bigg\{
\ln \left(2 \cosh{(\pi \Theta_q)} \right)
  \notag \\
&&
- \left( \frac{\beta |\Delta_0|}{2} \right)^2
\frac{\tanh{(\pi \Theta_q)}}{(\pi \Theta_q)}
\Bigg\}
\end{eqnarray*}
}

\noindent
where $\Theta_q = \left(\frac{\beta}{2\pi} \right)
\sqrt{x(1-x)v_F^2 \vec{q}^2 + |\Delta_0|^2}$.

\end{document}